\documentclass[12pt,english]{article}
\usepackage[T1]{fontenc} 
\usepackage{textcomp}
\usepackage[utf8]{inputenc} 
\usepackage{amsmath}
\usepackage{graphicx}
\usepackage{geometry}
\usepackage{slashed}
\usepackage{setspace}
\usepackage{bbm}
\makeatletter

\providecommand{\tabularnewline}{\\}

\newcommand{\GeV}{\ensuremath{\mathrm{GeV}}} 
 
\makeatother

\usepackage{babel}
\begin{document}
\title{Rare Exclusive Decays of the Z-boson into S-wave Quarkonia within the Bethe-Salpeter Formalism}

\author{
Asif Ali$^\text{a}$,
Yi-Jie Li$^\text{a}$,
Guang-Zhi Xu$^\text{a,1}$,
Kui-Yong Liu$^\text{b,a,1}$
\\[8pt] 
\parbox{\linewidth}{\small
$^\text{a}$ School of Physics, Liaoning University, Shenyang 110036, China\\
$^\text{b}$ School of Physics and Electronic Technology, Liaoning Normal University, Dalian 116029, China\\
$^\text{1}$ Corresponding authors: \texttt{xuguangzhi@lnu.edu.cn}, \texttt{liukuiyong@lnu.edu.cn}
}
}

\date{}

\maketitle

\begin{abstract}
This paper investigates the rare decays of the Z boson into S-wave quarkonia within the Bethe-Salpeter formalism. Both the production of double S-wave quarkonia and radiative decays into single S-wave quarkonium are analyzed. For double quarkonia production, i.e., $Z\to VV$ and $Z\to VP$ (where $V$ and $P$ represent vector and pseudoscalar quarkonia, respectively), we consider the leading order contribution from both QCD as well as electromagnetic transition via virtual photon (QED) amplitudes. The heavy quark limit is adopted to simplify calculations. Additionally, we have introduced another possible leading order channel for Z-boson decays to double S-wave quarkonia, where the Z boson decays into bottomonium plus charmonium via QED amplitude. Such processes may include $Z\to J/\psi+\Upsilon(1S)$, $Z\to\eta_{b}+J/\psi$ and $Z\to\eta_{c}+\Upsilon(1S)$. Moreover, we have also studied radiative Z boson decays to S-wave quarkonium, namely $Z\to X(Q\bar{Q})\gamma$, where $X(Q\bar{Q})=J/\psi$, $\Upsilon$, $\eta_{c}$, and $\eta_{b}$. Interestingly, for double charmonium and radiative charmonium production, our results are larger than the NRQCD finding, while for the bottomonium case, our findings are comparatively smaller. This shows that charmonium is a relativistic particle, while bottomonium is a non-relativistic particle.
\end{abstract}

\section{Introduction}\label{sec:introduction}

The study of heavy quarkonium production and decay in high-energy phenomena provides a remarkable testing ground for quantum chromodynamics (QCD), particularly in investigating the interplay between non-perturbative and perturbative dynamics. Moreover, such transitions also help us to understand the dynamics and structure of quarkonium. Among these processes, the rare decays of the Z boson to double quarkonia as well as the radiative decay of the Z boson to quarkonium are of particular importance because they offer unique insights into the interplay between electroweak and strong interaction dynamics. However, the Z boson predominantly decays into fermion-antifermion pairs; its decay into bound quarkonium states involves complex QCD effects, making it a valuable opportunity for understanding quarkonium production. In recent years, the decay of the Z boson to double quarkonium has been very actively investigated at the LHC. The CMS collaboration \cite{key-1} first reported upper limit and branching fraction for the Z boson decays to double vector quarkonia in 2018. Again, in 2022, they performed it and updated the upper limits for such rare decays at a 95\% confidence level \cite{key-2}. Moreover, experimental searches for the radiative Z-boson decays into heavy quarkonium have been conducted by ATLAS and CMS collaborations at the LHC \cite{key-3,key-4}, yet no significant signals have been detected. Over the past few years, several high-luminosity $e^+e^-$ colliders like ILC \cite{key-5}, Super Z factory \cite{key-6}, CEPC \cite{key-7}, and FCC-ee \cite{key-8} are set to run at the Z pole for a period of time; consequently, a significant yield of Z boson events is anticipated. We hope that they provide new opportunities to investigate rare Z boson decays.

Over the past several years, various theoretical approaches have been developed to examine these processes. These include non-relativistic quantum chromodynamics (NRQCD) and light-cone (LC) models. Each framework provides distinct insights into different energy scales of QCD: while potential models capture the leading non-relativistic behavior, the LC can incorporate relativistic and field-theoretical effects essential for accurately describing heavy quarkonium systems. In Ref.~\cite{key-9}, the authors calculated the branching ratio for leading-order (LO) QCD diagrams of decays $Z\to VV$ and $Z\to VP$. Similarly, in Ref.~\cite{key-10}, both QCD and QED diagrams are taken into account, and it turns out that the contributions from QED diagrams dominate over those from QCD diagrams. 
Moreover, in Ref.~\cite{key-11}, a next-to-leading-order (NLO) QCD analysis has been performed for the process $Z\to J/\psi J/\psi$. The QCD corrections produce a sizable enhancement of contributions from QCD diagrams by roughly a factor of 4 to 5, yet they substantially suppress the QED contributions. The total yield remains nearly identical to the calculation presented in Ref.~\cite{key-10}.
The Z-boson decays into double quarkonium (S-wave and P-wave) were studied in the LC formalism \cite{key-12}. In the same way, the Z-boson decays into a photon and a quarkonium have been studied in different models. In 1980, the first theoretical study for radiative decays of the Z boson into quarkonium was carried out by the authors of Ref.~\cite{key-13}, the authors there calculated the decay rates at LO in $\alpha_{s}$ and $v^2$. In Ref.~\cite{key-14}, the radiative decays of the Z boson into S-wave or P-wave charmonium were studied using both the NRQCD and light-cone distribution amplitude (LCDA) approaches. Similarly, the decays $Z\to J/\psi(\Upsilon)+\gamma$ were investigated in both the NRQCD and LCDA approaches in Ref.~\cite{key-15}, and they also calculated the indirect contribution where the Z-boson decays into a real photon and a virtual photon through a heavy-quark or W-boson loop, with the virtual photon fragmenting into a heavy quarkonium. Recently, the decay rates and branching fractions for the Z-boson decays into a photon plus S-wave were calculated up to $O(\alpha_{s}v^2)$ corrections \cite{key-16}. However, LO contribution is dominant in all processes mentioned above. 

Among existing models, the Bethe-Salpeter (BS) formalism \cite{key-17,key-18,key-19} stands out as one of the most powerful and covariant methods for describing relativistic bound states. As a non-perturbative approach rooted in quantum field theory, the BS equation provides a relativistic, dynamical framework for analyzing bound states. Its covariance allows it to describe internal quark motion and spin effects in a consistent manner, making it suitable for investigating hadronic properties over a wide energy range. It treats the meson as a two-body system of quark and antiquark interacting through an effective kernel that contains both one-gluon-exchange (OGE) and confining components. In the instantaneous approximation, the BS equation reduces to the Salpeter equations, whose solutions yield the meson mass spectrum and relativistic wave functions. These wave functions can then be used consistently to calculate observables such as decay constants, electromagnetic transition amplitudes, and decay widths, providing a unified theoretical framework consistent with quantum field theory principles.
For many years, heavy quarkonium production and decay have been studied using the BS model, yielding reliable results \cite{key-20,key-21,key-22,key-23,key-24}. 
In this work, we present the first calculation of rare Z-boson decays to S-wave quarkonia in the BS framework. Our results are systematically compared with NRQCD predictions, leading to some physically meaningful conclusions.

Our work is organized as follows. In Sec.~\ref{sec:bethe_salpeter_framework}, we present the BS formalism, including the derivation of basic equations and the numerical solution of Salpeter equations with a QCD-inspired quark-antiquark kernel to obtain the momentum-space radial wavefunctions for both vector (V) and pseudoscalar (P) S-wave quarkonium bound states. In Sec.~\ref{sec:decay_processes}, we systematically calculate the branching ratios for Z boson decay processes within the BS framework. We first analyze $Z\to VV$ decays (Sec.~\ref{subsec:z_to_vv}) and $Z\to VP$ decays (Sec.~\ref{subsec:z_to_pv}), accounting for both QCD and QED LO contributions for each channel. We then extend our analysis to a new class of LO QED-mediated decays where the Z boson decays into bottomonium plus charmonium (Sec.~\ref{subsec:z_to_bottomonium_charmonium_qed}). Additionally, we investigate the radiative decays $Z\to X(Q\bar{Q})+\gamma$ (with $X(Q\bar{Q}) = J/\psi$, $\Upsilon$, $\eta_{c}$, $\eta_{b}$) in Sec.~\ref{subsec:z_to_quarkonium_gamma}. Finally, we summarize our key findings and discuss the implications of relativistic effects in Sec.~\ref{sec:summary}.

\section{Bethe-Salpeter Framework }\label{sec:bethe_salpeter_framework}

\subsection{Instantaneous BS Equations for Heavy Quarkonium}\label{subsec:basic_equations}

Let us consider a quarkonium, comprising of a fermionic quark $Q$ and an antiquark $\bar{Q}$ of masses $m_{1}$ and $m_{2}$, respectively. The four-dimensional (4D) BS equation for $Q\bar{Q}$ system can be written as 
\begin{equation}
S^{-1}(p_{1})\Psi(P,q)S^{-1}(-p_{2})=i\int\frac{d^{4}k}{(2\pi)^{4}}V(P,q,k)\Psi(P,k)\label{eq:a}
\end{equation}

In the above equation, $\Psi(P,q)$ is a 4D-BS wave function and $V(P,q,k)$ is the internal kernel, responsible for interaction between quark and antiquark, while $S^{-1}(\pm p_{1,2})$ are their corresponding inverse propagators. Here, $p_{1},p_{2}$ are the momenta of the quark and antiquark, respectively, and $q$ is the internal relative momentum of the meson, which has mass $M$, and external momentum $P$. The external momentum and internal momentum are related to quark and antiquark momenta as
\[
p_{1,2\mu}=\hat{m}P_{\mu}\pm q_{\mu}.
\]

Here, $\hat{m}=\frac{1}{2}\left[1\pm\frac{(m_{1}^{2}-m_{2}^{2})}{M^{2}}\right]$, for equal masses $p_{1,2\mu}$ become
\[
p_{1,2\mu}=\frac{1}{2}P_{\mu}\pm q_{\mu}.
\]
While instantaneous approximation reduced the BS kernel to 3D by making it independent of the longitudinal component of $q$, it can be written as $V(P,q,k)=V(q_{\perp},k_{\perp})$. Where the 3D variable $q_{\perp\mu}$ is written $q_{\perp\mu}=q_{\mu}-\frac{q\cdot P}{M^{2}}P_{\mu}$, orthogonal to the meson momentum, i.e., $q_{\perp}\cdot P=0$ and the longitudinal component to the $P_{\mu}$ is $q_{\parallel}=\frac{q\cdot P}{M^{2}}P_{\mu}$. Therefore, we can write them in decomposed form as $q_{\mu}=(q_{\parallel},q_{\perp})$, where $q_{\parallel}$ acts as the time component. Correspondingly, we may have two Lorentz invariant variables:
\begin{align*}
q_{P}=P\cdot q/M & ,\quad q_{T}=\sqrt{q_{P}^{2}-q^{2}}=\sqrt{-q_{\perp}^{2}}.
\end{align*}

When $\vec{P}=0$, they become usual components $q^{0}$ and $|\vec{q}|$ respectively. 
Now, we can integrate over the longitudinal component of the 4D-BS volume element $\mathrm{d}^{4}k$ appearing on the right side of Eq.~\ref{eq:a} and obtain
\begin{equation}
S^{-1}(p_{1})\Psi(P,q)S^{-1}(-p_{2})=\int\frac{d^{3}k_{\perp}}{(2\pi)^{3}}V(q_{\perp},k_{\perp})\psi(k_{\perp})=\Gamma(q_{\perp}).\label{eq:c}
\end{equation}
Here,
\begin{equation}
\psi(k_{\perp})=\frac{i}{2\pi}\int\mathrm{d}k_{P}\Psi(P,k),\label{eq:b}
\end{equation}
and $\Gamma(q_{\perp})$ is the hadron-quark vertex function. Finally, the BS wavefunction reads as 
\begin{equation}
\Psi(P,q)=S_{1}(p_{1})\Gamma(q_{\perp})S_{2}(-p_{2})\label{eq:d}
\end{equation}

where, $S_{i}(\pm p_{i})$ is the fermionic propagator of the quarks, which can be decomposed as 
\begin{align*}
S_{i}(\pm p_{i}) & =\frac{\varLambda_{i}^{+}(q_{\perp})}{I(i)q_{p}+\frac{1}{2}M-\omega_{i}}+\frac{\varLambda_{i}^{-}(q_{\perp})}{I(i)q_{p}+\frac{1}{2}M+\omega_{i}}.
\end{align*}

Here we define
\begin{align}
\omega_{i}^{2}=m_{i}^{2}-q_{\perp}^{2}=m_{i}^{2}+\vec{q}^{2},\text{and }\varLambda_{i}^{\pm} & =\frac{1}{2\omega_{i}}[\frac{\slashed{P}}{M}\omega_{i}\pm I(i)(m_{i}+\slashed{q}_{\perp})]\label{eq:n1}
\end{align}

For the index assignments, $i=1,2$ correspond to quark and antiquark, respectively and $I(i)=(-1)^{i+1}$. In Eq.~\ref{eq:n1}, $\varLambda_{i}^{\pm}$ is energy projection operators, which operate on $\psi(q_{\perp})$ giving projected wave functions:
\begin{align*}
\psi^{\pm\pm}(q_{\perp}) & =\varLambda_{1}^{\pm}(q_{\perp})\frac{\slashed{P}}{M}\psi(q_{\perp})\frac{\slashed{P}}{M}\varLambda_{2}^{\mp}(q_{\perp}).
\end{align*}

We can use $\Gamma(q_{\perp})$to evaluate the transition amplitude. After performing contour integration over $q_{p}$ of Eq.~\ref{eq:d}, we obtain four independent Salpeter equations:
\begin{align}
(M-\omega_{1}-\omega_{2})\psi^{++}(q_{\perp}) & =\varLambda_{1}^{+}(q_{\perp})\Gamma(q_{\perp})\varLambda_{2}^{+}(q_{\perp})\nonumber \\
(M+\omega_{1}+\omega_{2})\psi^{--}(q_{\perp}) & =-\varLambda_{1}^{-}(q_{\perp})\Gamma(q_{\perp})\varLambda_{2}^{-}(q_{\perp})\label{eq:4inde}\\
\psi^{+-}(q_{\perp}) & =0\nonumber \\
\psi^{-+}(q_{\perp}) & =0.\nonumber 
\end{align}

And the normalization condition of the BS wavefunction is defined as 
\begin{equation}
\int\frac{d^{3}\vec{q}}{(2\pi)^{3}}Tr\left[\overline{\psi}^{++}(q_{\perp})\frac{\slashed{P}}{M}\psi^{++}(q_{\perp})\frac{\slashed{P}}{M}-\overline{\psi}^{--}(q_{\perp})\frac{\slashed{P}}{M}\psi^{--}(q_{\perp})\frac{\slashed{P}}{M}\right]=2P_{0}.\label{eq:normf}
\end{equation}
To solve the eigenvalue equation, we have to choose a definite kernel of the quark and antiquark in the bound state. It comprises two parts, a linear scalar interaction (confinement one) plus a vector interaction (single gluon exchange) \cite{key-20,key-37}, which can be written in the momentum space and in the rest frame of the bound state as
\begin{align}
V(\vec{q})= & V_{v}(\vec{q})+\gamma_{0}\otimes\gamma^{0}V_{s}(\vec{q})\nonumber \\
V_{v}(\vec{q})= & \frac{2\pi\alpha_{s}}{3\pi^{2}(\vec{q}^{2}+\beta^{2})}\nonumber \\
V_{s}(\vec{q})= & -\left(\frac{\lambda}{\beta}+V_{0}\right)\delta^{3}(\vec{q})+\frac{\lambda}{\pi^{2}}\frac{1}{(\vec{q}^{2}+\beta^{2})^{2}}\label{eq:ker}
\end{align}

Here, $\alpha_{s}(\vec{q}^{2})=\frac{12\pi}{(33-2N_{f})}\frac{1}{\ln(\mathbbm{e}+\vec{q}^{2}/\varLambda_{\text{QCD}}^{2})}$ is the running coupling of the one-loop QCD correction. Where $N_{f}=3$ for charmonium, $N_{f}=4$ for bottomonium; the constants $\lambda$, $\beta$, $\mathbbm{e}$, $V_{0}$ and $\Lambda_{\text{QCD}}$ are the parameters which define and characterize the potential. Now, we can evaluate the mass spectrum and radial wave function of vector and pseudoscalar quarkonium states.

\subsection{Mass spectral equation for Vector and Pseudoscalar Meson}\label{subsec:radial_wave_functions_mass_spectra}

The BS wave function for pseudoscalar ($J^{PC}=0^{-+}$ ) and vector ($J^{PC}=1^{--}$ ) are defined in Refs. \cite{key-22,key-23,key-19,key-37}.
We work within the heavy quark limit, i.e., all $1/m$ corrections are neglected \cite{key-23,key-19}.
\begin{align}
\psi_{p}(q_{\perp}) & =\phi_{p}(q_{\perp})[\slashed{P}+M_{p}]\gamma^5\label{eq:wevp}\\
\psi_{v}(q_{\perp}) & =\phi_{v}(q_{\perp})[\slashed{P}+M_{v}]\slashed{\epsilon}\nonumber 
\end{align}

In the above equation, $M_{v,p}$ is mass and $\phi_{v,p}(q_{\perp})$ is radial wavefunction of corresponding quarkonium, where $\epsilon$ denotes polarization vector of vector quarkonium. It is important to notice that, in Eq.~\ref{eq:4inde}, the positive wave $\psi^{++}(q_{\perp})$ function is dominant, and that the contribution of the negative wave function $\psi^{--}(q_{\perp})$ can be safely neglected. Since the mass of quark and antiquark is the same, we have taken $w_{1}=w_{2}=w$ and for sake of simplicity, we have used $V_{s}=V_{s}(q_{\perp},k_{\perp}),V_{v}=V_{v}(q_{\perp},k_{\perp})$.
By substituting functions Eq.~\ref{eq:wevp} into the first equations of Eq.~\ref{eq:4inde} and taking Dirac traces on both sides of the equations, we obtain an integral equation:
\begin{align}
(M_{v,p}-2w)\phi_{v,p}(q_{\perp}) & =\int\frac{d^{3}\vec{k}}{(2\pi)^{3}}(V_{v}-V_{s})\phi_{v,p}(k_{\perp}).\label{eq:eie}
\end{align}
However, under the heavy quark limit, i.e., $\omega=\sqrt{m^{2}-q^{2}_{\perp}}\sim m$,
Eq.~\ref{eq:eie} turns into 
\begin{align}
\phi_{v,p}(q_{\perp}) & =\frac{1}{E}\int\frac{\mathrm{d}^3\vec{k}}{(2\pi)^{3}}\bigl(V_{v}-V_{s}\bigr)\phi_{v,p}(k_{\perp}).\label{eq:eie-1}
\end{align}
Here $E=M_{v,p}-2m$ denotes the binding energy, and the corresponding
normalization condition determined by Eq.~\ref{eq:normf} reads
\begin{equation}
4M_{v,p}\int\frac{\mathrm{d}^3\vec{q}}{(2\pi)^{3}}\big|\phi_{v,p}(q_{\perp})\big|^{2}=1.
\end{equation}

In practical calculations, to obtain reliable BS wavefunctions and mass spectra,
we solve Eq.~\ref{eq:eie-1} numerically as an eigenvalue problem for a given interaction kernel. 
The radial wave functions of pseudoscalar and vector quarkonium states are presented in Fig.~1.

\begin{figure}[h]
\centering
\includegraphics[scale=0.6]{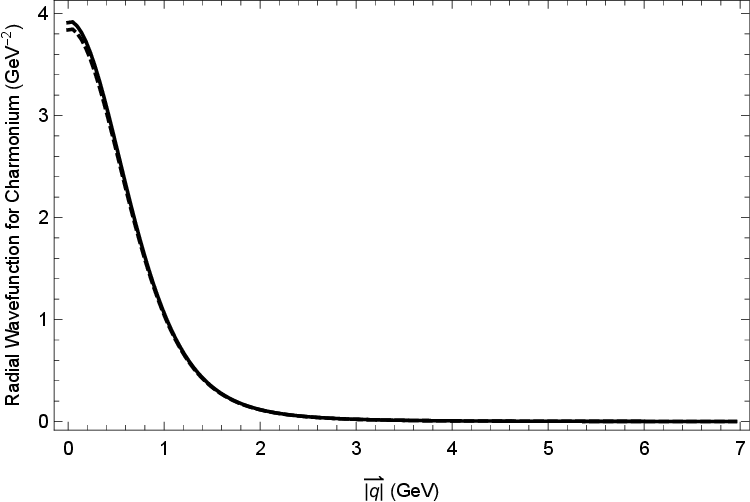} 
\includegraphics[scale=0.6]{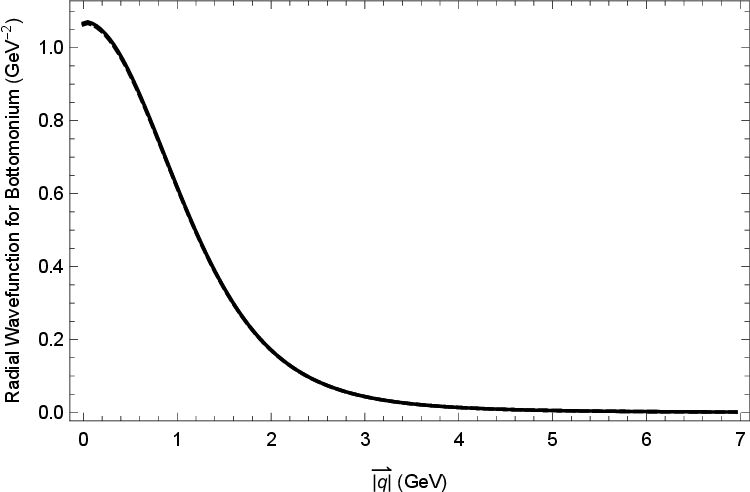}
\caption{The radial wave functions of the ground states ( $J^{PC}=0^{-+}$: Solid line, $J^{PC}=1^{--}$: dashed line) are displayed. The left one is for charmonium, and the right one is for bottomonium.}
\end{figure}

The fitting parameters for these states are shown in Table~\ref{tab:params}, which are taken from Ref.~\cite{key-37}, while $V_{0}$ is fixed by fitting the mass of the ground state of corresponding quarkonium.

\begin{table}[htbp]
\centering
\begin{tabular}{|c|c|c|c|c|c|}
\hline 
Quarkonium & $\mathbbm{e}$ & $\beta$ (\GeV) & $\lambda$ (\GeV$^2$) & $m$ (\GeV) & $\Lambda_{\text{QCD}}$ (\GeV)\\
\hline 
Charmonium & 2.7183 & 0.06 & 0.21 & 1.62 & 0.27\\
\hline 
Bottomonium & 2.7183 & 0.06 & 0.21 & 4.96 & 0.20\\
\hline 
\end{tabular}
\caption{Input parameters used in the present BS Model.}
\label{tab:params}
\end{table}

\section{Decay Processes}\label{sec:decay_processes}

\subsection{Z Boson Decays to a Pair of Vector Quarkonia, $Z\to VV$}\label{subsec:z_to_vv}

\begin{figure}[h]
\centering
\includegraphics[scale=0.6]{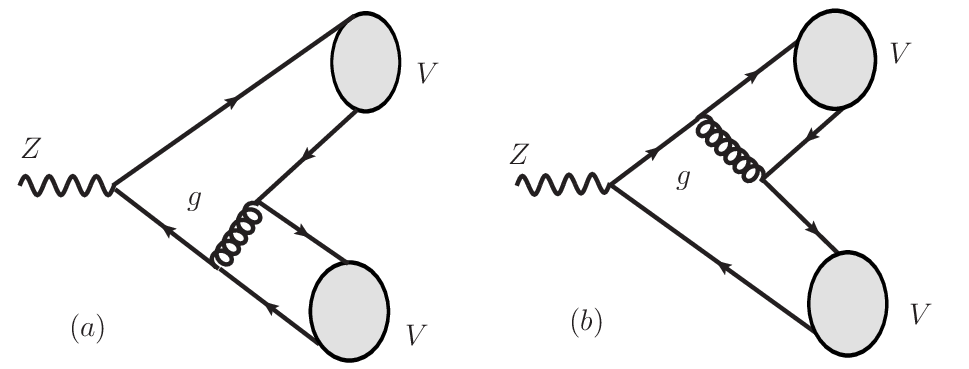}
\caption{Lowest-order QCD Feynman diagrams for the Z $\to$ $VV$ process. We can obtain two more diagrams by permuting these two diagrams.\label{fig:QCDZVV}}
\end{figure}

In this section, we investigated $Z\to VV$ process within the BS formalism. Let $P_{1},q_{1},\epsilon_{1}$ and $P_{2},q_{2},\epsilon_{2}$ denote the quarkonium momentum, quarks' relativistic momenta and the polarization vectors of the two outgoing vector quarkonia, respectively. Moreover, $\epsilon_{z}$ represents the polarization vector of the Z boson. We begin by evaluating QCD amplitude, which involves two types of lowest-order Feynman diagrams, shown in Fig.~\ref{fig:QCDZVV}, and the corresponding 4D-BS amplitude reads as:

\begin{multline}
\mathcal{M}_{\mathrm{QCD}}^{VV}=-i\frac{4C_{F}g_{s}^{2}gg_{a}^{Q}}{\cos\theta_{w}M_{\mathrm{Z}}^{2}}\int\frac{d^{4}q_{2}}{(2\pi)^{4}}\int\frac{d^{4}q_{1}}{(2\pi)^{4}}\biggl\{\mathrm{Tr}\biggl[\overline{\Psi}(P_{1},q_{1})\slashed{\varepsilon}_z\gamma^5\frac{(-\slashed{P}_{2}-\slashed{P}_{1}/2+m)}{((P_{2}+P_{1}/2)^{2}-m^{2})}\gamma^{\nu}\overline{\Psi}(P_{2},q_{2})\gamma^{\beta}+\\
\overline{\Psi}(P_{1},q_{1})\gamma^{\nu}\frac{(\slashed{P}_{1}+\slashed{P}_{2}/2+m)}{((P_{1}+P_{2}/2)^{2}-m^{2})}\slashed{\varepsilon}_z\gamma^5\overline{\Psi}(P_{2},q_{2})\gamma^{\beta}\biggr]\biggr\}g_{\nu\beta}\label{eq:zqcd1}
\end{multline}

where $\Psi(P_{1},q_{1})$ and $\Psi(P_{2},q_{2})$ are the relativistic 4D-BS wavefunction of the vector bosons and their corresponding conjugation is defined as; $\overline{\Psi}\equiv\gamma^{0}\Psi^{\dagger}\gamma^{0}$. $C_F = 4/3$ is the fundamental color factor of $\mathrm{SU}(3)$. Momenta of the quark and antiquark for both vector quarkonium are $\frac{1}{2}P_{1}\pm q_{1}$ and $\frac{1}{2}P_{2}\pm q_{2}$, respectively.

One can notice that the amplitude contains integration over the relative momentum of quarks and antiquark. Unfortunately, the propagators of gluon and quark may have explicitly depended on $q_{1}$ and $q_{2}$, which causes complicated calculations. However, employing the heavy quark approximation on the propagators \cite{key-23,key-24,key-19}, we have neglected the relative momentum in Eq.~\ref{eq:zqcd1}. 
The vertex factor for the $ZQ\bar{Q}$ interaction reads
$-i\frac{g}{2\cos\theta_{W}}\gamma^{\mu}\big(g_{v}^{Q}-g_{a}^{Q}\gamma^{5}\big)$.
The weak SU(2)$_{L}$ coupling constant is denoted by $g$, while $\theta_{W}$ represents the Weinberg angle. The axial-vector coupling of the $Z$ boson to a quark is given by:
\[
g_{a}^{Q}=T_{3}^{Q},
\]
where $T_{3}^{Q}$ corresponds to the third component of the quark's weak isospin. The strong interaction is characterized by the coupling constant $g_{s}$ ($\alpha_{s}=\frac{g_{s}^{2}}{4\pi}$), representing the strong fine-structure constant. It can be noticed that only $g_{a}^{Q}$ survives in the amplitude $\mathcal{M}_{\mathrm{QCD}}^{VV}$, because of the charge conjugate invariance. Using the notation of Salpeter wave function (Eq.~\ref{eq:b}), Eq.~\ref{eq:zqcd1} yields a 3D amplitude:
\begin{multline}
\mathcal{M}_{\mathrm{QCD}}^{VV}=i\frac{128\pi\alpha_{s}gg_{a}^{Q}}{3\cos\theta_{w}M_{\mathrm{Z}}^{4}}\int\frac{d^{3}\vec{q}_{1}}{(2\pi)^{3}}\int\frac{d^{3}\vec{q}_{2}}{(2\pi)^{3}}\biggl\{\mathrm{Tr}\biggl[\overline{\psi}(q_{\perp1})\slashed{\varepsilon}_z\gamma^5(-\slashed{P}_{2}-\slashed{P}_{1}/2+m)\gamma^{\nu}\overline{\psi}(q_{\perp2})\gamma^{\beta}+\\
\overline{\psi}(q_{\perp1})\gamma^{\nu}(\slashed{P}_{1}+\slashed{P}_{2}/2+m)\slashed{\varepsilon}_z\gamma^5\overline{\psi}(q_{\perp2})\gamma^{\beta}\biggr]\biggr\}g_{\nu\beta}\label{eq:zqcd2}
\end{multline}
Here, $\overline{\psi}(q_{\perp})$ is the conjugation of the 3B-BS wavefunction of vector quarkonium, defined in Eq.~\ref{eq:wevp}. After performing the Dirac trace of the previous equation, one can obtain:

\begin{equation}
\mathcal{M}_{\mathrm{QCD}}^{VV}=-\frac{2^{10}\pi g_{a}^{Q}g\alpha_{s}M_{v}^{2}}{3\cos\theta_{w}M_{\mathrm{Z}}^{4}}\int\frac{d^{3}\vec{q}_{1}}{(2\pi)^{3}}\int\frac{d^{3}\vec{q}_{2}}{(2\pi)^{3}}\phi_{1}(q_{\perp1})\phi_{2}(q_{\perp2})\epsilon_{\mu\nu\rho\sigma}\varepsilon_{z}^{\mu}\varepsilon_{1}^{\nu}\varepsilon_{2}^{\rho}(P_{1}-P_{2})^{\sigma}.\label{eq:qcdvam}
\end{equation}

Finally, squaring the amplitude, summing and averaging over the polarization of the final or initial particles, and using the traditional two-body decay formula, one can obtain the decay width:

\begin{equation}
\Gamma_{\mathrm{QCD}}^{VV}=\frac{2^{15}\pi}{27}\biggl(\frac{\alpha_{s}g_{a}^{Q}gM_{v}\xi_{V}^{2}}{\cos\theta_{w}M_{\mathrm{Z}}^{5}}\biggr)^{2}(M_{\mathrm{Z}}^{2}-4M_{v}^{2})^{5/2}\label{eq:de1}
\end{equation}

In the above expression, the factor from the exchange symmetry of identical particles has be considered. 
We introduce the quantity $\xi_{V}=\int\frac{d^{3}\vec{q}_{1}}{(2\pi)^{3}}\phi(q_{\perp})$, which can be computed numerically:
\begin{align}
\xi_{J/\psi} & =0.101GeV\nonumber \\
\xi_{\Upsilon(1S)} & =0.109GeV\label{eq:vfv}
\end{align}

There is another possible QED process $Z\to V+\gamma^*$, followed by the fragmentation of the off-shell photon $\gamma^*$ into $V$ (effectively a $Z\to V+V$ transition via an off-shell photon), a rare and theoretically interesting transition in the Standard Model. The Feynman diagram is shown in Fig.3, and the corresponding QED-BS amplitude can be written as:

\begin{multline}
\mathcal{M}_{\mathrm{QED}}^{VV}=i\frac{24\pi\alpha_{\text{em}}e_{Q}^{2}gg_{a}^{Q}}{\cos\theta_{w}M_{v}^{2}M_{\mathrm{Z}}^{2}}\int\frac{d^{3}\vec{q}_{1}}{(2\pi)^{3}}\int\frac{d^{3}\vec{q}_{2}}{(2\pi)^{3}}\biggl\{\mathrm{Tr}\biggl[\overline{\psi}(q_{\perp1})\gamma^{\beta}\biggr]\cdot\mathrm{Tr}\biggl[\overline{\psi}(q_{\perp2})\slashed{\varepsilon}_z\gamma^5(-\slashed{P}_{1}-\slashed{P}_{2}/2+m)\gamma^{\nu}\biggr]\\
+\mathrm{Tr}\biggl[\overline{\psi}(q_{\perp2})\gamma^{\beta}\biggr]\cdot\mathrm{Tr}\biggl[\overline{\psi}(q_{\perp2})\slashed{\varepsilon}_z\gamma^5(-\slashed{P}_{2}-\slashed{P}_{1}/2+m)\gamma^{\nu}\biggr]\biggr\}g_{\nu\beta}\label{qedvv}
\end{multline}

\begin{figure}[h]
\centering
\includegraphics[scale=0.6]{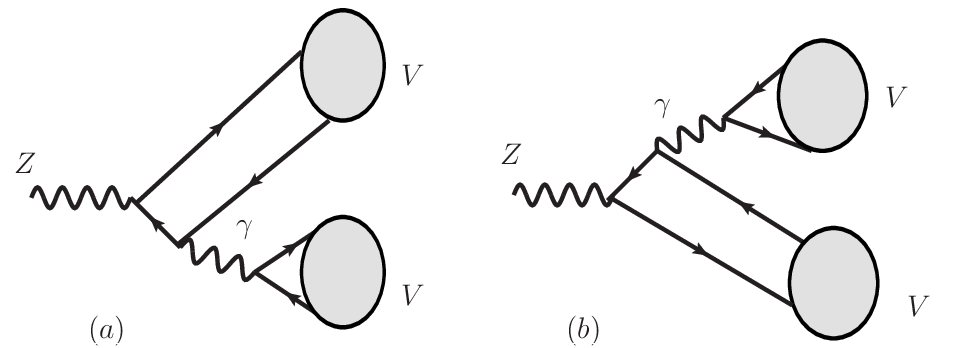}
\caption{Lowest-order QED Feynman diagrams for the Z $\to$ $VV$ process. We can obtain two more diagrams by permuting these two diagrams.}
\end{figure}

In the above expressions, the $M_{v}^{2}$ in the denominator comes from the photon propagator, and $e_{Q}$ denotes the electric charge of the corresponding quark. Similar to the QCD contribution, $g_{v}^{Q}$ vanishes here too. After evaluating Dirac trace we obtain:
\begin{equation}
\mathcal{M}_{\mathrm{QED}}^{VV}=\frac{3\times2^{7}\pi gg_{a}^{Q}\alpha_{\text{em}}e_{Q}^{2}}{\cos\theta_{w}M_{\mathrm{Z}}^{2}}\int\frac{d^{3}\vec{q}_{1}}{(2\pi)^{3}}\int\frac{d^{3}\vec{q}_{2}}{(2\pi)^{3}}\phi_{1}(q_{\perp1})\phi_{2}(q_{\perp2})\epsilon_{\mu\nu\rho\sigma}\varepsilon_{z}^{\mu}\varepsilon_{1}^{\nu}\varepsilon_{2}^{\rho}(P_{1}-P_{2})^{\sigma}\label{eq:qedvam}
\end{equation}

Generally, it seems like $\mathcal{M}_{\mathrm{QED}}^{VV}$ will be suppressed because of electromagnetic coupling. However, one can see that the photon propagator gives the factor $1/M_{v}^{2}$, which is comparatively larger than the factor of $4/M_{\mathrm{Z}}$, arising from the gluon propagator in the QCD amplitude. As a result the QED amplitude is larger than QCD, and it can be easily computed by dividing both amplitudes. After some straightforward mathematical steps, we get:
\begin{equation}
\Gamma_{\mathrm{QED}}^{VV}=3\times2^{9}\pi\biggl(\frac{e_{Q}^{2}\alpha_{\text{em}}g_{a}^{Q}g\xi_{V}^{2}}{\cos\theta_{w}M_{v}M_{\mathrm{Z}}^{3}}\biggr)^{2}(M_{\mathrm{Z}}^{2}-4M_{v}^{2})^{5/2}\label{eq:qedvv1}
\end{equation}

We have used $\alpha_{s}(M_{\mathrm{Z}})=0.118$, $\alpha_{\text{em}}(M_{J/\psi})=1/132.64$ and $\alpha_{\text{em}}(M_{\Upsilon(1S)})=1/131.87$, which were also used in Ref.~\cite{key-10}, and the branching fraction can be calculated by adapting $\Gamma/\Gamma_{t}$. Here, $\Gamma_{t}=2.4995\ \GeV$ is the total decay width of the Z boson. We have taken the mass of Z boson and other constants from Particle Data Group (PDG) and we also determined the value of the mass of vector quarkonium $M_{v}$ from Eq.~\ref{eq:eie}, which is approximately equal to experimental values. Table~\ref {tab:vv_br} presents the branching fractions calculated within the BS and NRQCD frameworks.

\begin{table}[h]
\centering
\begin{tabular}{|c|c|c|c|c|}
\hline 
Process & BS(our) & NRQCD\cite{key-10} & \cite{key-11} & LHC(upper lim)\cite{key-2}\tabularnewline
\hline 
\hline 
(QCD)$Z\to J/\psi J/\psi$ & $4.57\times10^{-13}$ & $1.5\times10^{-13}$ &  & $11\text{\texttimes}10^{\text{\textminus}7}$\tabularnewline
\hline 
(QED+QCD)$Z\to J/\psi J/\psi$ & $3.26\times10^{-10}$ & $1.1\times10^{-10}$ & $1.1\times10^{-10}$ & \tabularnewline
\hline 
(QCD$)Z\to\Upsilon(1S)\Upsilon(1S)$ & $5.24\times10^{-12}$ & $6.8\times10^{-12}$ &  & $1.8\text{\texttimes}10^{\text{\textminus}6}$\tabularnewline
\hline 
(QED+QCD)$Z\to\Upsilon(1S)\Upsilon(1S)$ & $3.36\times10^{-13}$ & $4.4\times10^{-13}$ &  & \tabularnewline
\hline 
\end{tabular}
\caption{Branching fractions of Z decays into double vector quarkonia are calculated
in BS model, as well as the recent NRQCD findings and the recently
updated upper limit at LHC. ``(QCD)'' denotes the contribution from pure QCD processes only, while ``(QED+QCD)'' includes both QED, QCD contributions and their interference terms.\label{tab:vv_br}}
\end{table}

We find that the branching fraction for charmonium is enhanced by an order of $10^{3}$, due to the QED amplitude contribution, while for bottomonium, there is a slight difference. This is due to larger mass and smaller charge of bottom the quark. One can observe that the amplitudes of QCD and QED have opposite signs, leading to destructive interference. Another important point is that the findings for charmonium are comparatively larger in the BS formalism, while for bottomonium, NRQCD gives comparatively larger values.

\subsection{Z Boson Decays to Pseudoscalar--Vector Quarkonium Pairs, $Z\to PV$ }\label{subsec:z_to_pv}

Similarly, rare decays of $Z\to PV$ can be investigated, and the BS wavefunction for pseudoscalar is defined in Eq.~\ref{eq:wevp}. This process resembles the remarkable double charmonium production process in electron-positron annihilation, $e^{-}+e^{+}\to\eta_{c}+J/\psi$, and the lowest-order QCD diagram for this transition is the same as shown in Fig.2, but replace one of the vectors by pseudoscalar boson. This process also receives contribution from both QED and QCD amplitudes. The color-singlet 3D-BS QCD amplitude is written as:
\[
\mathcal{M}_{\mathrm{QCD}}^{VP}=-\frac{2^{11}\pi g_{v}^{Q}g\alpha_{s}M_{v}}{3\cos\theta_{w}M_{\mathrm{Z}}^{4}}\int\frac{d^{3}\vec{q}_{1}}{(2\pi)^{3}}\int\frac{d^{3}\vec{q}_{2}}{(2\pi)^{3}}\phi_{1}(q_{\perp1})\phi_{2}(q_{\perp2})\epsilon_{\mu\nu\rho\sigma}\varepsilon_{z}^{\mu}\varepsilon_{1}^{\nu}P_{1}^{\rho}P_{2}^{\sigma}
\]

In the above expression, $\phi_{1}(q_{\perp1})$ and $\phi_{2}(q_{\perp2})$ are the radial wave functions of vector and pseudoscalar charmonium, respectively. In this transition $g_{v}^{Q}=T_{3}^{Q}-2e_{Q}\sin^{2}\theta_{w}$, survives, which is responsible for the Z boson to quark-antiquark pair vector vertex coupling. For brevity, we set $M_{p}=M_{v}$ as an approximation. Following traditional mathematical steps, the corresponding decay width is:
\[
\Gamma_{\mathrm{QCD}}^{VP}=\frac{2^{17}\pi}{27}\biggl(\frac{\alpha_{s}gg_{v}^{Q}M_{v}\xi_{V}\xi_{P}}{\cos\theta_{w}M_{\mathrm{Z}}^{4}}\biggr)^{2}(M_{\mathrm{Z}}^{2}-4M_{v}^{2})^{3/2}.
\]

The numerical values of $\xi_{V}$ for Vector quarkonium are already given in Eq.~\ref{eq:vfv}, and those for pseudoscalar ($\xi_{P}$) are given by:
\begin{align*}
\xi_{\eta_{c}} & =0.103\ \GeV\\
\xi_{\eta_{b}} & =0.110\ \GeV
\end{align*}

Like the $Z\to VV$ case, we can evaluate the QED contribution to this process. The Feynman diagrams are the same as in Fig.3. However, only two diagrams are possible because angular conservation does not allow a photon to decay into a pseudoscalar meson. The 3D-BS amplitude can be written as:
\[
\mathcal{M}_{\mathrm{QED}}^{VP}=\frac{3\times2^{7}\pi g_{v}^{Q}g\alpha_{\text{em}}e_{Q}^{2}}{3\cos\theta_{w}M_{\mathrm{Z}}^{2}M_{v}}\int\frac{d^{3}\vec{q}_{1}}{(2\pi)^{3}}\int\frac{d^{3}\vec{q}_{2}}{(2\pi)^{3}}\phi_{1}(q_{\perp1})\phi_{2}(q_{\perp2})\epsilon_{\mu\nu\rho\sigma}\varepsilon_{z}^{\mu}\varepsilon_{1}^{\nu}P_{1}^{\rho}P_{2}^{\sigma}.
\]

The corresponding decay width is:
\begin{equation}
\Gamma_{\mathrm{QED}}^{VP}=3\times2^{9}\pi\biggl(\frac{e_{Q}^{2}\alpha_{\text{em}}gg_{v}^{Q}\xi_{V}\xi_{P}}{\cos\theta_{w}M_{v}M_{\mathrm{Z}}^{3}}\biggr)^{2}(M_{\mathrm{Z}}^{2}-4M_{v}^{2})^{3/2}\label{eq:dpvqed}
\end{equation}

The corresponding branching fraction values of our model and recent NRQCD findings are presented in Table~\ref{tab:vp_br}.

\begin{table}[htbp]
\centering
\begin{tabular}{|c|c|c|}
\hline 
Process & BS(our) & NRQCD \cite{key-10}\\
\hline
\hline 
(QCD)$Z\to J/\psi\eta_{c}$ & $2.80\times10^{-13}$ & $9.1\times10^{-14}$\tabularnewline
\hline 
(QED+QCD)$Z\to J/\psi\eta_{c}$ & $4.63\times10^{-11}$ & $1.5\times10^{-11}$\tabularnewline
\hline 
(QCD)$Z\to\Upsilon(1S)\eta_{b}$ & $1.07\times10^{-11}$ & $4.9\times10^{-11}$\tabularnewline
\hline 
(QED+QCD)$Z\to\Upsilon(1S)\eta_{b}$ & $4.19\times10^{-12}$ & $1.9\times10^{-11}$\tabularnewline
\hline 
\end{tabular}
\caption{Branching fractions for the process $Z\to\Upsilon(1S)+\eta_{b}$ and $Z\to J/\psi+\eta_{c}$ are shown, calculated in BS model alongside recently obtained NRQCD results. Here, ``(QCD)'' denotes the contribution from pure QCD processes only, while ``(QED+QCD)'' includes both QED, QCD contributions and their interference terms. }
\label{tab:vp_br}
\end{table}

Similar to the $Z\to VV$ process, the QED contributions are significant here. Moreover, here as well we observe destructive interference due to the opposite signs of QED and QCD amplitudes. Furthermore, while NRQCD predictions are larger for bottomonium production, our results are comparatively larger for charmonium. 

\subsection{Z Boson Decays to Bottomonium and Charmonium }\label{subsec:z_to_bottomonium_charmonium_qed} 

We have introduced another possible LO channel for Z boson decay to double S-wave quarkonia, where Z boson decays into bottomonium plus charmonium via QED amplitude. The Feynman diagram is shown in Fig.3. The decay processes are $Z\to J/\psi+\Upsilon(1S)$, $Z\to\eta_{b}+J/\psi$, and $Z\to\eta_{c}+\Upsilon(1S)$. We investigate them one by one.

Let's begin with $Z\to J/\psi+\Upsilon(1S)$, the Feynman diagram is similar to Fig.3, just replace a charmonium vector with a bottomonium vector, and the corresponding 3D-BS amplitude is:
\begin{equation}
\begin{split}
\mathcal{M}_{Z\to J/\psi+\Upsilon(1S)} &= 3\frac{2^{6}\pi gg_{a}^{Q}\alpha_{\text{em}}e_{b}e_{c}M_{J/\psi}M_{\Upsilon(1S)}}{\cos\theta_{w}} \\
&\quad \int\frac{d^{3}\vec{q}_{1}}{(2\pi)^{3}}\int\frac{d^{3}\vec{q}_{2}}{(2\pi)^{3}}\phi_{J/\psi}(q_{\perp1})\phi_{\Upsilon(1S)}(q_{\perp2})\biggl[d_{1}\epsilon_{\mu\nu\rho\sigma}\varepsilon_{z}^{\mu}\varepsilon_{1}^{\nu}\varepsilon_{2}^{\rho}P^{\sigma}-d_{2}\epsilon_{\mu\nu\rho\sigma}\varepsilon_{z}^{\mu}\varepsilon_{1}^{\nu}\varepsilon_{2}^{\rho}P_{2}^{\sigma}\biggr]
\end{split}
\label{eq:MV_two_lines}
\end{equation}

In the above expression, $\varepsilon_{1}$ and $\varepsilon_{2}$ are the polarization vectors, and $\phi_{J/\psi}(q_{\perp1})$ and $\phi_{\Upsilon(1S)}(q_{\perp2})$ are the radial wave functions of vector charmonium and vector bottomonium, respectively. For brevity, we have set $\alpha_{\text{em}}(M_{\Upsilon(1S)})=\alpha_{\text{em}}(M_{J/\psi})$ as an approximation. This transition proceeds via $g_{a}^{Q}$ and the factors $d_{1}$ and $d_{2}$ are defined below:
\begin{align*}
d_{1} & =\frac{1}{\frac{1}{2}(M_{J/\psi}^{2}-M_{\Upsilon(1S)}^{2}+M_{\mathrm{Z}}^{2})M_{J/\psi}^{2}}\\
d_{2} & =\frac{1}{\frac{1}{2}(M_{\Upsilon(1S)}^{2}-M_{J/\psi}^{2}+M_{\mathrm{Z}}^{2})M_{\Upsilon(1S)}^{2}}
\end{align*}

Since both outgoing vectors have different masses, the magnitude of the 3-momentum of each outgoing particle is read as:
\[
|\vec{P}|=\frac{1}{2M_{\mathrm{Z}}}\sqrt{\left[M_{\mathrm{Z}}^{2}-(M_{J/\psi}+M_{\Upsilon(1S)})^{2}\right]\left[M_{\mathrm{Z}}^{2}-(M_{J/\psi}-M_{\Upsilon(1S)})^{2}\right]}.
\]

Finally, we are ready to present the decay width expression:
\begin{equation}
\Gamma_{Z\to J/\psi+\Upsilon(1S)}=3\times2^{9}\pi\biggl(\frac{e_{c}e_{b}\alpha_{\text{em}}gg_{a}^{Q}\xi_{J/\psi}\xi_{\Upsilon(1S)}}{\cos\theta_{w}}\biggr)^{2}A_{Vv}\label{eq:qedVvD}
\end{equation}

For brevity, we have introduced $A_{Vv}$ in Eq.~\ref{eq:qedVvD}, one can find the full expression in the Appendix. It is interesting to note that we can obtain Eq.~\ref{eq:qedvv1} by setting $M_{J/\psi}=M_{\Upsilon(1S)}=M_{v}$, $\xi_{J/\psi}=\xi_{\Upsilon(1S)}=\xi_{V}$, and $e_{c}=e_{b}=e_{Q}$ in Eq.~\ref{eq:qedVvD}, the only difference is extra factor of $\frac{1}{2}$ (quantum statistical factor). The corresponding numerical value of the branching fraction is given in Table~\ref{tab:mix_br}.

Now we will analyze $Z\to\eta_{b}+J/\psi$ and $Z\to\eta_{c}+\Upsilon(1S)$. Like $Z\to PV$, they also have only two LO Feynman diagrams and $Z\to\eta_{b}+J/\psi$ decay via $g_{v}^{b}$ and $Z\to\eta_{c}+\Upsilon(1S)$ decay via $g_{v}^{c}$. Here, $g_{v}^{c}$ and $g_{v}^{c}$ denotes Z-boson to $(c\bar{c})$ and Z boson to $(b\bar{b})$ vector vertex coupling, respectively. The corresponding 3D-BS amplitude for both processes reads as:
\begin{equation}
\mathcal{M}_{Z\to\eta_{b}+J/\psi}=-\frac{3\times2^{6}\pi gg_{v}^{b}\alpha_{\text{em}}e_{b}e_{c}M_{J/\psi}}{\cos\theta_{w}}l_{1}\int\frac{d^{3}\vec{q}_{1}}{(2\pi)^{3}}\int\frac{d^{3}\vec{q}_{2}}{(2\pi)^{3}}\phi_{J/\psi}(q_{\perp1})\phi_{\eta_b}(q_{\perp2})\epsilon_{\mu\nu\rho\sigma}\varepsilon_{z}^{\mu}\varepsilon_{1}^{\nu}P_{\eta_{b}}^{\rho}P_{J/\psi}^{\sigma}\label{eq:vcpb}
\end{equation}
\begin{equation}
\mathcal{M}_{Z\to\eta_{c}+\Upsilon(1S)}=-\frac{3\times2^{6}\pi gg_{v}^{c}\alpha_{\text{em}}e_{b}e_{c}M_{\Upsilon(1S)}}{\cos\theta_{w}}l_{2}\int\frac{d^{3}\vec{q}_{1}}{(2\pi)^{3}}\int\frac{d^{3}\vec{q}_{2}}{(2\pi)^{3}}\phi_{\eta_{c}}(q_{\perp1})\phi_{\Upsilon(1S)}(q_{\perp2})\epsilon_{\mu\nu\rho\sigma}\varepsilon_{z}^{\mu}\varepsilon_{2}^{\nu}P_{\eta_{c}}^{\rho}P_{\Upsilon(1S)}^{\sigma}\label{eq:vbcp}
\end{equation}

In Eq.~\ref{eq:vcpb} and Eq.~\ref{eq:vbcp}, $\varepsilon_{1}$ and $\varepsilon_{2}$ are the polarization vectors of charmonium and bottomonium, respectively. The factors $l_{1}$ and $l_{2}$ are given below:
\begin{align*}
l_{1} & =\frac{1}{\frac{1}{2}(M_{J/\psi}^{2}-M_{\eta_{b}}^{2}+M_{\mathrm{Z}}^{2})M_{J/\psi}^{2}}\\
l_{2} & =\frac{1}{\frac{1}{2}(M_{\Upsilon(1S)}^{2}-M_{\eta_{c}}^{2}+M_{\mathrm{Z}}^{2})M_{\Upsilon(1S)}^{2}}
\end{align*}

One can find expressions for decay widths for these processes using previous procedure:
\begin{equation}
\Gamma_{Z\to\eta_{b}+J/\psi}=3\times2^{9}\pi\biggl(\frac{e_{c}e_{b}\alpha_{\text{em}}gg_{v}^{b}\xi_{J/\psi}\xi_{\eta_{b}}}{\cos\theta_{w}}\biggr)^{2}A_{1}\label{eq:devcpb}
\end{equation}
\begin{equation}
\Gamma_{Z\to\eta_{c}+\Upsilon(1S)}=3\times2^{9}\pi\biggl(\frac{e_{c}e_{b}\alpha_{\text{em}}gg_{v}^{c}\xi_{\eta_{c}}\xi_{\Upsilon(1S)}}{\cos\theta_{w}}\biggr)^{2}A_{2}\label{eq:dvbpc}
\end{equation}

For simplicity, $A_1$ and $A_2$ are introduced in Eqs.~\ref{eq:devcpb} and \ref{eq:dvbpc}, respectively, with their full expressions given in the Appendix. We can also reproduce Eq.~\ref{eq:dpvqed} by replacing the corresponding parameters of the bottomonium with those of the charmonium or vice versa. The branching fractions for both processes are also given in Table~\ref{tab:mix_br}.

\begin{table}[h]
\centering
\begin{tabular}{|c|c|c|c|}
\hline
Process & BS(our) & NRQCD(LO) & NRQCD($\alpha_s,v^2$)\cite{Wang:2026qws} \\
\hline
$Z\rightarrow J/\psi+\Upsilon(1S)$ & $1.1\times10^{-10}$ & $8.69\times10^{-11}$ & $1.43\times10^{-12}$\\
\hline 
$Z\rightarrow\eta_{b}+J/\psi$ & $5.0\times10^{-11}$ & $3.7\times10^{-11}$ &\\
\hline 
$Z\rightarrow\eta_{c}+\Upsilon(1S)$ & $1.6\times10^{-12}$ & 1.12$\times10^{-12}$ &\\
\hline
\end{tabular}
\caption{Branching fractions for the processes $Z \to J/\psi + \Upsilon(1S)$, $Z \to \eta_b + J/\psi$, and $Z \to \eta_c + \Upsilon(1S)$, calculated within the BS model alongside NRQCD predictions. For the NRQCD(LO) results, we adopt the input matrix elements and quark masses: $\langle O^{J/\psi}(^3S_1)\rangle=1.2\,\mathrm{GeV}^3$, $\langle O^{\Upsilon}(^3S_1)\rangle=10.9\,\mathrm{GeV}^3$, $m_c=1.5\,\mathrm{GeV}$, $m_b=4.7\,\mathrm{GeV}$. All remaining parameters are consistent with those adopted throughout this work.
 }
\label{tab:mix_br}
\end{table}

As expected, branching fraction value of $Z\to J/\psi+\Upsilon(1S)$ lies between that of QED amplitude of $Z\to J/\psi+J/\psi$ and $Z\to\Upsilon(1S)+\Upsilon(1S)$. Similarly, branching fraction value of $Z\to\eta_{b}+J/\psi$ lies between that of $Z\to J/\psi+\eta_{c}$ and $Z\to\Upsilon(1S)+\eta_{b}$. However, the value for $Z\to\eta_{c}+\Upsilon(1S)$ process does not obey that pattern, and it is comparatively smaller. The reason is that $Z$ decays through the vector coupling $g_{v}^{c}$, which is weaker than $g_{v}^{b}$, and a virtual photon can easily decay into lighter particles ($M_{J/\psi}$) than into massive ones ($M_{\Upsilon(1S)}$).
Our results agree with LO NRQCD predictions in order of magnitude. Remarkably, the NLO radiative and relativistic corrections both provide large negative contributions to this process, reducing the $\alpha_s v^2$ predictions by two orders of magnitude.

\subsection{Radiative Z Boson Decays to Quarkonium States }\label{subsec:z_to_quarkonium_gamma}

\begin{figure}[h]
\centering
\includegraphics[scale=0.6]{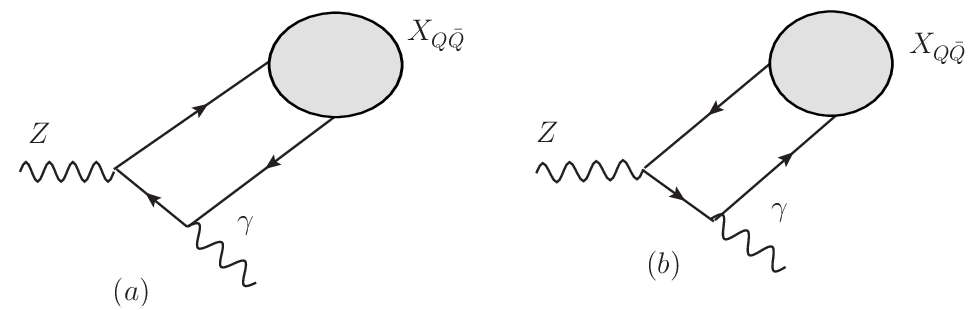}
\caption{Lowest-order Feynman diagrams for $Z\to X(Q\bar{Q})\gamma$ process.\label{fig:radiativedecays}}
\end{figure}

The rare radiative decay of the Z boson into a quarkonium state plus a photon, $Z\to X(Q\bar{Q})\gamma$, provides a valuable probe into non-perturbative aspects of QCD. In the Standard Model, such decays proceed primarily through secondary production mechanisms, where the Z boson decays to a pair of heavy quarks, which then radiatively bind into the final quarkonium state. The general 3D-BS amplitude corresponding to Fig.~\ref{fig:radiativedecays} can be read as,

\begin{multline}
\mathcal{M}_{Z\to X\gamma}=-\frac{\sqrt{N_c} ee_{Q}g}{2\cos\theta_{w}}\int\frac{d^{3}\vec{q}}{(2\pi)^{3}}\Biggl\{\mathrm{Tr}\biggl[\overline{\psi}(q_{\perp})\Bigl(\slashed{\varepsilon}_z(g_{v}^{Q}-g_{a}^{Q}\gamma^5)\frac{(-\slashed{k}-\slashed{P}/2+m)}{((k+P/2)^{2}-m^{2})}\slashed{\epsilon}_{\gamma} \\
+\slashed{\epsilon}_{\gamma}\frac{(\slashed{k}+\slashed{P}/2+m)}{((k+P/2)^{2}-m^{2})}\slashed{\varepsilon}_z(g_{v}^{Q}-g_{a}^{Q}\gamma^5)\Bigr)\biggr]\Biggr\}\label{eq:amgma}
\end{multline}

Here $\psi(q_{\perp})$ is the 3D-BS wavefunction of vector or pseudoscalar quarkonium, and $k$ and $\epsilon_{\gamma}$ denote momentum and polarization vector of radiative photon. After substituting the BS wavefunction of the corresponding quarkonium in Eq.~\ref{eq:amgma} and taking the Dirac trace, we obtained their amplitudes:
\begin{align}
\mathcal{M}_{Z\to P\gamma}= & -i\frac{2^{3}\sqrt{3}ee_{Q}gg_{v}}{\cos\theta_{w}(M_{\mathrm{Z}}^{2}-M_{p}^{2})}\int\frac{d^{3}\vec{q}}{(2\pi)^{3}}\phi_{p}(q_{\perp})\epsilon_{\mu\nu\rho\sigma}k^\mu P_1^\nu \varepsilon_z^\rho \varepsilon_\gamma^\sigma\nonumber \\
\mathcal{M}_{Z\to V\gamma}= & -i\frac{2^{3}\sqrt{3}ee_{Q}gg_{a}M_{v}}{\cos\theta_{w}(M_{\mathrm{Z}}^{2}-M_{v}^{2})}\int\frac{d^{3}\vec{q}}{(2\pi)^{3}}\phi_{v}(q_{\perp})\epsilon_{\mu\nu\rho\sigma}k^\mu \varepsilon_z^\nu \varepsilon_1^\rho \varepsilon_\gamma^\sigma\label{Mgamvp}
\end{align}

Due to charge-conjugate invariance, $\mathcal{M}_{Z\to P\gamma}$ proceed via $g_{v}^{Q}$ and $\mathcal{M}_{Z\to V\gamma}$ via $g_{a}^{Q}$. One can define the two-body radiative decay width for $Z\to X(Q\bar{Q})+\gamma$ as follows;
\begin{equation}
\Gamma_{Z\to X(Q\bar{Q})\gamma}=\frac{1}{3}\cdot\frac{1}{2M_{\mathrm{Z}}}\cdot\frac{M_{\mathrm{Z}}^{2}-M_{X(Q\bar{Q})}^{2}}{8\pi M_{\mathrm{Z}}^{2}}\cdot|\mathcal{M}|^{2}\label{def}
\end{equation}

Here, $M_{X(Q\bar{Q})}$ and $\mathcal{M}$ represent mass and amplitude of the corresponding quarkonium. In order to determine the analytical decay width of the pseudoscalar and vector quarkonium, we substituted Eq.~\ref{Mgamvp} in Eq.~\ref{def} and performed straightforward mathematical steps:
\begin{align*}
\Gamma_{Z\to P\gamma} & =\frac{2}{\pi}\biggl(\frac{ee_{Q}gg_{v}^{Q}\xi_{p}}{\cos\theta_{w}}\biggr)^{2}\frac{(M_{\mathrm{Z}}^{2}-M_{p}^{2})}{M_{\mathrm{Z}}^{3}}\\
\Gamma_{Z\to V\gamma} & =\frac{2}{\pi}\biggl(\frac{ee_{Q}gg_{a}^{Q}\xi_{V}}{\cos\theta_{w}}\biggr)^{2}\frac{(M_{\mathrm{Z}}^{4}-M_{v}^{4})}{M_{\mathrm{Z}}^{5}}
\end{align*}

Finally, the branching fractions can be calculated by following the usual steps and are shown in Table~\ref{tab:gamma_br}.

\begin{table}[htbp]
\centering
\begin{tabular}{|c|c|c|c|}
\hline 
Process & BS(our) & NRQCD($\alpha_sv^2$) \cite{key-16} & Light Cone \cite{key-14} \\
\hline
\hline
$Z\to J/\psi\gamma$ & $1.6\times10^{-7}$ & $1.04\times10^{-7}$ & $8.8\times10^{-8}$\tabularnewline
\hline 
$Z\to\eta_{c}\gamma$ & $2.4\times10^{-8}$ & $1.32\times10^{-8}$ & $9.4\times10^{-9}$\tabularnewline
\hline 
$Z\to\Upsilon(1S)\gamma$ & $4.6\times10^{-8}$ & $5.4\times10^{-8}$ & \tabularnewline
\hline 
$Z\to\eta_{b}\gamma$ & $2.24\times10^{-8}$ & $2.8\times10^{-8}$ & \tabularnewline
\hline 
\end{tabular}
\caption{Branching fractions for the process $Z\to\eta_{c}(\eta_{b})+\gamma$ and $Z\to J/\psi(\Upsilon)+\gamma$ calculated in BS model, along with recently calculated results of NRQCD and and those from the light cone approach.}
\label{tab:gamma_br}
\end{table}

One can notice in Table~\ref{tab:gamma_br} that the values for the vector are comparatively larger than those for the pseudoscalar quarkonium. The main reason is that the production of $\eta_{c}(\eta_{b})+\gamma$ is through the vector coupling, while the production of $J/\psi(\Upsilon)+\gamma$ is through the axial-vector coupling of the Z boson, and the axial-vector coupling is stronger than the vector coupling in these transitions. Moreover, similar to double quarkonia production, the BS findings are larger for charmonium, and the NRQCD findings are greater for bottomonium. 

\section{Summary }\label{sec:summary}

For the first time, we employ the BS formalism to investigate the rare decays of the Z boson into S-wave quarkonia. These decay modes include double S-wave charmonia, double S-wave bottomonia, associated S-wave charmonium-bottomonium pairs, and the radiative decays of the Z boson into S-wave quarkonia. Currently, these decays are actively investigated both experimentally and theoretically. Our primary focus is on non-perturbative effects, so we use the BS model, which is well-suited for describing such effects, as relativistic effects play an important role in hadronization, particularly for charmonium. For the double quarkonia decay modes, we have considered the amplitudes of both QCD and QED channels, and found that the QED amplitude makes a significant contribution. However, destructive interference occurs due to the opposite signs of these two amplitudes. The total branching ratios are therefore far below the current experimental limits recently reported by the CMS Collaboration \cite{key-2}. For the Z boson decay modes to bottomonium plus charmonium via the QED amplitude alone, the calculated branching fraction of $Z\to J/\psi+\Upsilon(1S)$ lies, as expected, between those for the QED-only processes $Z\to J/\psi+J/\psi$ and $Z\to\Upsilon(1S)+\Upsilon(1S)$. Following the same trend, the branching fraction of $Z\to\eta_{b}+J/\psi$ lies between those of $Z\to J/\psi+\eta_{c}$ and $Z\to\Upsilon(1S)+\eta_{b}$. In contrast, the branching fraction for $Z\to\eta_{c}+\Upsilon(1S)$ is considerably smaller. When it comes to Z boson radiative decays, our results for charmonium channels are consistent with the latest upper limits set by the LHC Collaboration \cite{key-3}.

Additionally, it is noteworthy that charmonium production yields larger values within the BS formalism, whereas bottomonium production gives larger values within NRQCD. This indicates that charmonium is inherently relativistic, while bottomonium is a non-relativistic particle. This picture is consistent with the fact that the BS model is a covariant approach that naturally incorporates relativistic corrections, whereas NRQCD is formulated within a non-relativistic framework. Further evidence supporting this conclusion comes from calculations of the cross section for the process $e^+e^-\to J/\psi+\eta_{c}$ at B factories. In Refs. \cite{key-23,key-24,key-39}, the authors computed the cross section for the LO amplitude using the BS model, obtaining values of $22.3$~fb, $20.7$~fb, and $21.75$~fb, respectively. These values fall between the experimental results from Belle ($25.6$~fb) \cite{key-25,key-26} and BaBar ($17.6$~fb) \cite{key-27}. In contrast, the LO contribution within NRQCD is far too small \cite{key-28,key-29}. To resolve this discrepancy, substantial higher-order corrections (either $\alpha_{s}$ \cite{key-30,key-31} or relativistic corrections\cite{key-32,key-33,key-34}) are required, which amount to nearly twice the LO value \cite{key-29}. Recently, theoretical calculations for this process have been extended to the $\alpha_s^2$ order, providing a more precise theoretical basis \cite{key-35,key-36}.
An important point to note is that the only difference between the BS predictions within the heavy quark limit (as calculated in Ref.\cite{key-23} and in our work) and
LO NRQCD results for both $e^{+}e^{-}\rightarrow J/\psi+\eta_{c}$
and rare Z boson decays into S-wave quarkonia lies in the non-perturbative
part (radial wavefunction). 
For charmonium systems, the quark--antiquark relative velocity squared takes a typical value $v^{2}\approx0.3$, which renders relativistic corrections prominent. As a fully covariant framework, the BS formalism may be thus far more appropriate for charmonium observables than NRQCD. For bottomonium, however, $v^{2}\approx0.1$, so the non-relativistic NRQCD expansion becomes the preferable approach.

Rare Z boson decays could nevertheless provide a promising probe for new physics, and we hope that further searches for these decays will be investigated at upcoming high-luminosity facilities, such as the FCC-ee, ILC, CEPC, and Super Z factory.

\section*{Acknowledgements}\label{sec:acknowledgements}
The authors would like to express their gratitude to Dr. Xiao-Peng Wang, who performed the LO NRQCD calculations for associated charmonium--bottomonium decay processes. This work was supported by the National Natural Science Foundation of China (No. 11705078, 12575087).

\section*{Appendix }\label{sec:appendix}

Below we present the complete expressions for $A_{Vv}$, $A_{1}$, and $A_{2}$, which are defined in Eqs.~\ref{eq:qedVvD}, \ref{eq:devcpb}, and \ref{eq:dvbpc} for the corresponding charmonium--bottomonium decay processes.

\begin{flalign*}
A_{Vv} &= \frac{\sqrt{\left[M_{\mathrm{Z}}^{2}-\left(M_{\Upsilon(\mathrm{1S})}-M_{J/\psi}\right)^{2}\right]\left[M_{\mathrm{Z}}^{2}-\left(M_{\Upsilon(\mathrm{1S})}+M_{J/\psi}\right)^{2}\right]}}
{M_{\Upsilon(\mathrm{1S})}^{2}M_{J/\psi}^{2}M_{\mathrm{Z}}^{5}\left[M_{\mathrm{Z}}^{4}-\left(M_{\Upsilon(\mathrm{1S})}^{2}-M_{J/\psi}^{2}\right)^{2}\right]^{2}} \times \\
&\quad\biggl[ 
    \left(M_{\Upsilon(\mathrm{1S})}^{2}-M_{J/\psi}^{2}\right)^{6} +
    \left(M_{\Upsilon(\mathrm{1S})}^{2}-M_{J/\psi}^{2}\right)^{4}\left(M_{\Upsilon(\mathrm{1S})}^{2}+M_{J/\psi}^{2}\right)M_{\mathrm{Z}}^{2} \\
&\quad -2\left(M_{\Upsilon(\mathrm{1S})}^{2}-M_{J/\psi}^{2}\right)^{2}\left(M_{\Upsilon(\mathrm{1S})}^{4}-10M_{\Upsilon(\mathrm{1S})}^{2}M_{J/\psi}^{2}+M_{J/\psi}^{4}\right)M_{\mathrm{Z}}^{4} \\
&\quad -2\left(M_{\Upsilon(\mathrm{1S})}^{2}+M_{J/\psi}^{2}\right)\left(M_{\Upsilon(\mathrm{1S})}^{4}-10M_{\Upsilon(\mathrm{1S})}^{2}M_{J/\psi}^{2}+M_{J/\psi}^{4}\right)M_{\mathrm{Z}}^{6} \\
&\quad +\left(M_{\Upsilon(\mathrm{1S})}^{4}-18M_{\Upsilon(\mathrm{1S})}^{2}M_{J/\psi}^{2}+M_{J/\psi}^{4}\right)M_{\mathrm{Z}}^{8} +
    \left(M_{\Upsilon(\mathrm{1S})}^{2}+M_{J/\psi}^{2}\right)M_{\mathrm{Z}}^{10}
\biggr] &
\end{flalign*}

\begin{flalign*}
A_{1} &= \frac{\left[\left(M_{\mathrm{Z}}^{2}-(M_{\eta_{b}}-M_{J/\psi})^{2}\right)\left(M_{\mathrm{Z}}^{2}-(M_{\eta_{b}}+M_{J/\psi})^{2}\right)\right]^{3/2}}
{M_{J/\psi}^{2}M_{\mathrm{Z}}^{3}\left(M_{J/\psi}^{2}-M_{\eta_{b}}^{2}+M_{\mathrm{Z}}^{2}\right)^{2}} &
\end{flalign*}

\begin{flalign*}
A_{2} &= \frac{\left[\left(M_{\mathrm{Z}}^{2}-(M_{\Upsilon(\mathrm{1S})}-M_{\eta_{c}})^{2}\right)\left(M_{\mathrm{Z}}^{2}-(M_{\Upsilon(\mathrm{1S})}+M_{\eta_{c}})^{2}\right)\right]^{3/2}}
{M_{\Upsilon(\mathrm{1S})}^{2}M_{\mathrm{Z}}^{3}\left(M_{\Upsilon(\mathrm{1S})}^{2}-M_{\eta_{c}}^{2}+M_{\mathrm{Z}}^{2}\right)^{2}} &
\end{flalign*}

\end{document}